\documentclass[preprint,showkeys,secnumarabic,showpacs,amsmath,amssymb,superscriptaddress,]{revtex4}
\usepackage{graphicx}
\begin{document}
\title{Constant-roll Inflation in Brane Induced Gravity Cosmology}

\author{A. Ravanpak}
\email{a.ravanpak@vru.ac.ir}
\affiliation{Department of Physics, Vali-e-Asr University of Rafsanjan, Rafsanjan, Iran}

\author{G. F. Fadakar}
\email{g.farpour@vru.ac.ir}
\affiliation{Department of Physics, Vali-e-Asr University of Rafsanjan, Rafsanjan, Iran}

\date{\small {\today}}

\begin{abstract}

In this article we study a constant-roll inflationary model in the context of the DGP brane-world cosmology caused by a quintessence scalar field. We determine an analytical solution for the Friedman equation coupled to the equation of motion of the scaler field. The evolution of the primordial scalar and tensor perturbations is also studied. To check the viability of the model we use numerical approaches and plot some figures. Our results for the scalar spectral index and the tensor to scaler ratio show good consistency with observations for given values of model parameters.

\end{abstract}



\maketitle

\section{Introduction}

Inflation is a very short period at early stages of the Universe nearly just after the Big-Bang in which the Universe experienced a huge rapidly accelerated expansion. It first proposed in \cite{Starobinsky2}-\cite{Guth}, to solve some of the long-standing problems of the hot big-bang model of cosmology, such as the flatness problem, the horizon problem and the monopole problem. In addition, the inflationary epoch provides some inhomogeneities in the Universe arisen from vacuum fluctuations. So, it can explain the large scale structure of the Universe and the measured anisotropies in the cosmic microwave background radiation \cite{Liddle}-\cite{Hinshaw}. Usually, an isolated standard scalar field called inflaton, drives inflation. If the inflaton field rolls down its potential as much slow as it supports a long-enough period of inflation, the inflationary scenario called slow-roll inflation \cite{Linde},\cite{Albrecht}. One can find lots of articles about the slow-roll inflationary paradigm in the literature \cite{Faraoni}-\cite{Ravanpak}.

Other inflationary solutions in which one can ignore the assumption of the inflaton slow-roll, were proposed and studied in \cite{Starobinsky}-\cite{Namjoo}. Among them, in \cite{Tsamis}-\cite{Namjoo}, the authors have investigated the so called ultra-slow-roll inflation. After the generalization of this scenario in \cite{Martin}, the idea of a constant-roll inflation was raised and developed in \cite{Motohashi}-\cite{Setare}. Although viable models of constant-roll inflation appear to be close to the slow-roll ones, but it is still useful in calculating small corrections to the slow-roll results \cite{Motohashi}. One of the most interesting properties of the constant-roll inflationary model is the prediction of non-Gaussianities \cite{Namjoo}-\cite{Motohashi2},\cite{Cai}.

Independent of the above subjects, extra dimensional theories of gravity which have come out of the string theory have recently attracted a great deal of attention. Several important five dimensional (5D) cosmological models have been proposed to explain the weakness of gravity and the hierarchy problem \cite{Arkani}-\cite{Randall2}. These models consider our four dimensional (4D) Universe as a brane embedded in a 5D space-time named bulk. Based on these brane-world models, the standard model of particle physics is confined to the brane, while gravity can propagate into the extra dimension. One interesting way to generalize the gravitational action of a 5D theory is to add an induced gravity correction term through considering a 4D scalar curvature in addition to the matter Lagrangian within the brane action. A well-known example of brane induced gravity scenario is the so called DGP model \cite{Dvali}. In this brane-world model, the 5D bulk is considered to be infinite and Minkowski. Also, the bulk cosmological constant and the brane tension is assumed to be zero. This model consists of two solutions distinguished with a parameter $\epsilon=\pm1$. The case $\epsilon=+1$, named self-accelerating branch, explains the late-time acceleration of the Universe without need to a dark energy component, though it suffers from the ghost problem. On the opposite side, the case $\epsilon=-1$, called normal branch needs a dark energy component to explain the late-time acceleration but it is ghost free. The dynamics of the Universe in a DGP scenario has been frequently studied in the literature \cite{Yin}-\cite{Wu}. Specifically, the idea of inflation has also been investigated in the context of the basic DGP brane-world cosmology and its generalizations \cite{Cai2}-\cite{Ravanpak7}.

Herein, we will study the idea of a constant-roll inflation in the context of the DGP brane-world model. One of the main motivations to choose a brane context is the latest results of Planck satellite according which the so called D-brane inflation is one of the best inflationary models that is in good agreement with observations \cite{Akrami}. We will consider a quintessence scalar field as the inflaton field and we will obtain the evolutionary equations of the model and their solutions. Then, we will study the evolution of the primordial inhomogeneities. To show the viability of the model, we will use a few figures to compare our results with observations. The structure of the article is as following: in Sec.2, we introduce the model, obtain the equations and solve them, analytically. Sec.3, deals with the perturbation theory. In this section we calculate the most important perturbation parameters of the model. Numerical discussions present in Sec.4 in which we try to show the consistency between our results and observations. Finally, a summary and the conclusions are presented in Sec.5.

\section{constant-roll inflation in DGP cosmology}

The general action of the DGP brane induced gravity model is written as
\begin{equation}\label{Friedmann}
  S=\frac{1}{2\kappa_5^2}\int d^5 x\sqrt{-g_5}R_5 +\int d^4 x\sqrt{-g_4}\cal{ L}
\end{equation}
in which the first term corresponds to the Einstein-Hilbert action in a 5D Minkowski bulk and the second one shows the contribution of the induced gravity localized
on the brane. Also, $R_5$ and $\kappa_5$, is the 5D Ricci scalar and gravitational constant, respectively. In addition, we have considered the effective 4D Lagrangian on the brane as
\begin{equation}\label{L}
{\cal L}= {\frac{\gamma}{2\kappa_4^2}R_4+{\cal{L}}_m}
\end{equation}
The parameter $\gamma$ is a dimensionless constant parameter which controls the strength of the induced gravity term and satisfies the condition $0\leq\gamma\leq1$. The case $\gamma=1$ gives the basic DGP model while there is no brane induced gravity term when $\gamma=0$. Also, $R_4$ is the 4D Ricci scalar, ${\cal{L}}_m$ is the matter Lagrangian on the brane and $\kappa_4^2=8\pi/M_4^2$ in which $M_4$ is the 4D Planck mass. In fact, our action is a special case of a more general induced gravity one in which the cosmological constant of the bulk and the tension of the brane have been set to zero.
Considering the spatially flat Friedmann-Robertson-Walker metric on the brane, we obtain the Friedmann equation of the model as below
\begin{equation}\label{Friedmann2}
    H^2=\frac{\kappa_4^2}{3\gamma}\left(\sqrt{\rho+\frac{\rho_0}{2}}+\epsilon \sqrt{\frac{\rho_0}{2}}\right)^2
\end{equation}
in which $\rho$ is the energy density of the matter content of the Universe and $\rho_0 =\frac{6\kappa_4^2}{\gamma\kappa_5^4}$. Since inflation happens in the very early Universe, we impose the high energy condition $\rho\gg \rho_0$ [65],[66]. Under this condition the above equation reduces to
\begin{equation}\label{Friedmann3}
    H^2=\frac{\kappa_4^2}{3\gamma}\left(\sqrt{\rho}+\epsilon \sqrt{\frac{\rho_0}{2}}\right)^2
\end{equation}
In the following we will focus on the normal branch $\epsilon = -1$.

Assuming the Universe is exclusively filled by a homogeneous scalar field $\phi$ as the inflaton field, with energy density $\rho=\dot\phi^2/2+V(\phi)$ in which $V$ stands for the inflaton potential we obtain the following effective Friedmann equation
\begin{equation}\label{Friedmann5}
  H^2+4\kappa_4\sqrt{\frac{\rho_0}{24\gamma}}H=\frac{\kappa_4^2}{3\gamma}{\rho}- {\frac{\kappa_4^2}{6\gamma}\rho_0}
\end{equation}
and also
\begin{equation}\label{Raychaudhuri}
   \dot H (2H+4\kappa_4\sqrt{\frac{\rho_0}{24\gamma}})=-\frac{\kappa_4^2}{\gamma}H\dot\phi^2
\end{equation}

Moreover, the condition of constant-roll inflation is expressed as
\begin{equation}\label{cr}
\ddot\phi+\beta H\dot\phi=0
\end{equation}
It is clear that the standard slow-roll inflation occurs when $\beta=0$ while the ultra-slow-roll case corresponds to $\beta=3$. Since the constant-roll inflation interpolates between these two limiting cases, so we expect $0\leq\beta\leq3$. But, using cosmological observational data and according to the results in \cite{Motohashi2}, it lies in the small interval $0.01\leq\beta\leq0.02$, approximately. Considering the relation $a=e^N$, one can easily find the following solution for the above equation
\begin{equation}\label{dphi}
\dot\phi=\dot\phi_0e^{-\beta N}
\end{equation}
in which $N$ indicates the number of $e$-folds and $\dot\phi_0$ is a constant of integration.
If we rewrite Eq.(\ref{Raychaudhuri}) as
\begin{equation}\label{H}
\frac{dH}{dN}(2H+4\kappa_4\sqrt{\frac{\rho_0}{24\gamma}})=-\frac{\kappa_4^2}{\gamma}\dot\phi^2
\end{equation}
and replace $\dot\phi$ with Eq.(\ref{dphi}) then a simple integration with respect to $N$ leads to the following equation
\begin{equation}\label{H1}
H^2+4\kappa_4\sqrt{\frac{\rho_0}{24\gamma}}H=\frac{\kappa_4^2}{2\beta\gamma}\dot\phi_0^2e^{-2\beta N}+\cal C
\end{equation}
in which $\cal C$, is a constant of integration. Comparing with Eq.(\ref{Friedmann5}), one can obtain the inflaton potential as below
\begin{equation}\label{V}
V(N)=\frac{3-\beta}{2\beta}\dot\phi_0^2e^{-2\beta N}+V_0
\end{equation}
where $V_0=\frac{3\gamma}{\kappa_4^2}{\cal {C}}+\frac{\rho_0}{2}$. Assuming $V(N)$ is the only responsible of the entire evolution of the inflaton field including the graceful exit, reheating and so on, one can consider $V_0$, as a late-time cosmological constant which in turn satisfies the constraint $V_0\ll V$ during inflation. Because the inflaton field needs a potential with $dV/dN<0$, we conclude again the constraint $\beta<3$. To find $V$ in terms of $\phi$, we first rewrite Eq.(\ref{dphi}) as
\begin{equation}\label{N}
\dot\phi=H\frac{d\phi}{dN}=\dot\phi_0e^{-\beta N}
\end{equation}
Then using Eq.(\ref{H1}) we obtain
\begin{equation}\label{phiN}
\phi(N)=\sqrt{\frac{3\gamma}{\kappa_4^2}}\int{\dot\phi_0\frac{e^{-\beta N}dN}{{\sqrt{\frac{3}{2\beta}\dot\phi_0^2 e^{-2\beta N}+V_0}}-\sqrt{\frac{\rho_{0}}{2}}}}
\end{equation}
This equation has an analytic solution as below if we ignore $V_0$ compared to the other term in the square root in the denominator (Regarding to the prior discussions about $V_0$):
\begin{equation}\label{p}
\phi-\phi_0=-\sqrt{\frac{2\gamma}{\beta\kappa_4^2}}\ln\left(\sqrt{\frac{3}{2\beta}\dot\phi_0^2}e^{-\beta N}-\sqrt{\frac{\rho_0}{2}}\right)
\end{equation}
in which $\phi_0$ is an integrating constant. Then, one comes quickly to the following relation
\begin{equation}\label{N}
e^{-2\beta N}=\frac
{2\beta}{3\dot\phi_0^2}\left(e^{\sqrt{\frac{\beta\kappa_4^2}{2\gamma}}(\phi_0-\phi)}+\sqrt{\frac{\rho_0}{2}}\right)^2
\end{equation}
Replacing Eq.(\ref{N}) in Eq.(\ref{V}) we find
\begin{equation}\label{Vphi}
V(\phi)=\frac
{3-\beta}{3}\left(e^{\sqrt{\frac{\beta\kappa_4^2}{2\gamma}}(\phi_0-\phi)}+\sqrt{\frac{\rho_0}{2}}\right)^2
\end{equation}
Also, using Eqs.(\ref{Friedmann3}), (\ref{dphi}) and (\ref{N}) one can obtain $H$ in terms of $\phi$ as below
\begin{equation}\label{Hphi}
H=\sqrt{\frac{\kappa_4^2}{3\gamma}}e^{\sqrt{\frac{\beta\kappa_4^2}{2\gamma}}(\phi_0-\phi)}.
\end{equation}

Meanwhile, one of the most important parameter in studying inflationary models is the first slow-roll parameter $\varepsilon=-\frac{\dot H}{H^2}$. Using Eqs.(\ref{dphi}), (\ref{N}) and (\ref{Hphi}) in our model, we have
\begin{equation}\label{e}
\varepsilon=\frac{\beta\left(e^{\sqrt{\frac{\beta\kappa_4^2}{2\gamma}}(\phi_0-\phi)}+\sqrt{\frac{\rho_0}{2}}\right)}{e^{\sqrt{\frac{\beta\kappa_4^2}{2\gamma}}(\phi_0-\phi)}}
\end{equation}
Before entrance to the next section we need to obtain $\varepsilon$ in terms of the number of $e$-folds $N$. To this aim, we first try to calculate $N$ versus $\phi$. Again, using Eqs.(\ref{dphi}), (\ref{N}) and (\ref{Hphi}), we find:
\begin{equation}\label{number}
N=\int Hdt=\int_\phi^{\phi_{e}} \frac{H}{\dot\phi}d\phi=\frac{1}{\beta}\ln\left[\frac{e^{\sqrt{\frac{\beta\kappa_4^2}{2\gamma}}(\phi_0-\phi)}+\sqrt{\frac{\rho_0}{2}}}{e^{\sqrt{\frac{\beta\kappa_4^2}{2\gamma}}(\phi_0-\phi_e)}+\sqrt{\frac{\rho_0}{2}}}\right]
\end{equation}
Using Eq.(\ref{e}) and considering the condition $\varepsilon=1$ that describes the end of inflation one can easily obtain
\begin{equation}\label{phiend}
e^{\sqrt{\frac{\beta\kappa_4^2}{2\gamma}}(\phi_0-\phi_e)}=\frac{\beta}{1-\beta}\sqrt{\frac{\rho_0}{2}}
\end{equation}
Substituting the above result into Eq.(\ref{number}) we realize that
\begin{equation}\label{numberphi}
N=\frac{1}{\beta}\ln\left[\frac{(1-\beta)\left[e^{\sqrt{\frac{\beta\kappa_4^2}{2\gamma}}(\phi_0-\phi)}+\sqrt{\frac{\rho_0}{2}}\right]}{\sqrt{\frac{\rho_0}{2}}}\right]
\end{equation}
Finally, using Eqs.(\ref{e}) and (\ref{numberphi}), we find the important relation
\begin{equation}\label{eN}
\varepsilon=\frac{\left(\frac{\beta}{1-\beta}\right)e^{\beta N}}{\left(\frac{1}{1-\beta}\right)e^{\beta N}-1}
\end{equation}

\section{inflation perturbations}

In this section, we explore the linear perturbation theory in inflation utilizing a linearly perturbed Friedmann-Robertson-Walker metric which comprises both scalar and tensor perturbations. We calculate the power spectrum of scalar and tensor perturbations, ${\cal P_{\cal R}}$ and ${\cal P}_{g}$, which are useful quantities in characterizing the properties of perturbations.

The scalar perturbation can be obtained using Mukhanov-Sasaki equation \cite{Mukhanov},\cite{Sasaki}. Since the scalar field only exists on the brane, we use the same perturbation equation as in the standard 4D cosmology, that is
\begin{equation}\label{MS}
v_k''+\left(k^2-\frac{z''}{z}\right)v_k=0
\end{equation}
in which $z=\frac{a\dot\phi}{H}$, $v_k$ is the mode function that is related to the curvature perturbation $\zeta$ via $\kappa_4v_k=z\zeta_k$, $k$ stands for the wave number and prime means derivative with respect to the conformal time $\tau$. We follow the procedure in \cite{Motohashi}, \cite{Yi} and \cite{Gao2}. Using the definition of slow-roll parameters $\varepsilon=-\frac{\dot H}{H^2}$ and $\eta=-\frac{\ddot\phi}{H\dot\phi}$, since in the constant-roll scenario $\eta=\beta=constant$, we get to the first order approximation of $\varepsilon$:
\begin{equation}\label{z}
\frac{z''}{z}=\frac{1}{\tau^2}\left(\nu^2-\frac{1}{4}\right), \quad \nu^2=\frac{9}{4}-3\beta+\beta^2+6\varepsilon-8\varepsilon\beta+2\varepsilon\beta^2
\end{equation}
Since $\nu$ is a constant the solution of Eq.(\ref{MS}) can be given as the Hankel function of order $\nu$ as below:
\begin{equation}\label{mode}
v_k=\frac{\sqrt\pi}{2}\exp{\left[i\left(\nu+\frac{1}{2}\right)\frac{\pi}{2}\right]}\sqrt{-\tau}H_\nu^{(1)}(-k\tau)
\end{equation}
Therefore, the power spectrum of the scalar perturbation, ${\cal P_{\cal R}}=\frac{k^3}{2\pi^2}|\zeta_k^2|$, on super horizon scales $k\ll aH$, becomes
\begin{equation}\label{PR}
{\cal P_{\cal R}}=2^{2\nu-4}\left(\frac{\kappa_4^2}{\varepsilon}\right)\left[\frac{\Gamma(\nu)}{\Gamma(3/2)}\right]^2\left(\frac{H}{2\pi}\right)^2\left(1-\frac{\varepsilon}{1+2\beta}\right)^{2\nu-1}\left(\frac{k}{aH}\right)^{3-2\nu}
\end{equation}
in which we have used the asymptotic formula $\lim_{x\rightarrow0}H_\nu^{(1)}(x)\simeq\frac{-i}{\pi}\Gamma(\nu)(\frac{x}{2})^{-\nu}$ \cite{Motohashi}.

On the other hand, the tensor perturbations that produce gravitational waves during inflation, play an important role in the present model because of the leakage of gravitons into the bulk. Using the junction conditions at the brane, the effect of the extra dimension emerges as a change in the normalization of zero-mode metric fluctuations on the brane \cite{Bouhmadi2}. In the constant-roll inflation up to leading order approximation of $\varepsilon$, the second slow-roll parameter $\eta$, does not appear in the tensor perturbation equations. Therefore we use here the form of the amplitude of tensor perturbations derived in \cite{Bouhmadi2}. Then, using the notation of \cite{Lidsey} the amplitude of tensor perturbations in our model can be expressed as
\begin{equation}\label{pg}
{\cal P}_{g}=8\kappa_4^2\left(\frac{H}{2\pi}\right)^2F_\gamma^2(x)
\end{equation}
in which
\begin{equation}\label{Fx}
F_\gamma^{-2}(x)=\gamma+(1-\gamma)\left[{\sqrt{(1+x^2)}-x^2\arcsin\frac{1}{x}}\right]
\end{equation}
Here, $x=H/\bar{\mu}$, and $\bar{\mu}$ represents the energy scale associated with the bulk curvature or the anti de Sitter scale. Since in the DGP scenario the 5D bulk is infinite and Minkowski one can utilize the approximation $H\gg\bar{\mu}$, according with $F_\gamma^2(H/\bar{\mu})\approx1/\gamma$ (\cite{Bouhmadi2}).

Another useful quantity in perturbation theory is the ratio of the power spectrum of tensor perturbations and scalar perturbations called tensor-to-scalar ratio which defined as
\begin{equation}\label{r}
r=\frac{{\cal P}_{g}}{\cal P_{\cal R}}
\end{equation}
that in our model reduces to
\begin{equation}\label{r2}
r\approx2^{7-2\nu}\left(\frac{\varepsilon}{\gamma}\right)\left[\frac{\Gamma(3/2)}{\Gamma(\nu)}\right]^2\left(1-\frac{\varepsilon}{1+2\beta}\right)^{1-2\nu}
\end{equation}

There are two other important parameters related to the scalar perturbations, the scalar spectral index $n_s-1=d\ln{\cal P_{\cal R}}/d\ln k$ which describes the slope of the power spectrum $\cal P_{\cal R}$, and the running in the scalar spectral index parameter $n_{run}=dn_s/d\ln k$ which shows the scale dependence of primordial scalar fluctuations. Here, $k$ stands for the wave number and the interval in the wave number is related to the number of $e$-folds through $d\ln k=-dN$. In our model they become:
\begin{equation}\label{ns}
n_s-1=3-2\nu
\end{equation}
and
\begin{equation}\label{nrun}
n_{run}=\frac{1}{\nu}\frac{d\nu^2}{dN}
\end{equation}
\section{discussions}

With attention to Eqs.(\ref{r2}) and (\ref{z}), it is obvious that the trajectories in $(r-n_s)$ plane depend on both the constant-roll parameter $\beta$ and the induced parameter $\gamma$, while considering Eqs.(\ref{nrun}), (\ref{z}) and (\ref{eN}), one can easily conclude that the trajectories in $(n_{run}-n_s)$ plane only depend on $\beta$ and are insensitive with respect to the changes in $\gamma$.

\begin{figure}[h]
\centering
\includegraphics[width=8cm]{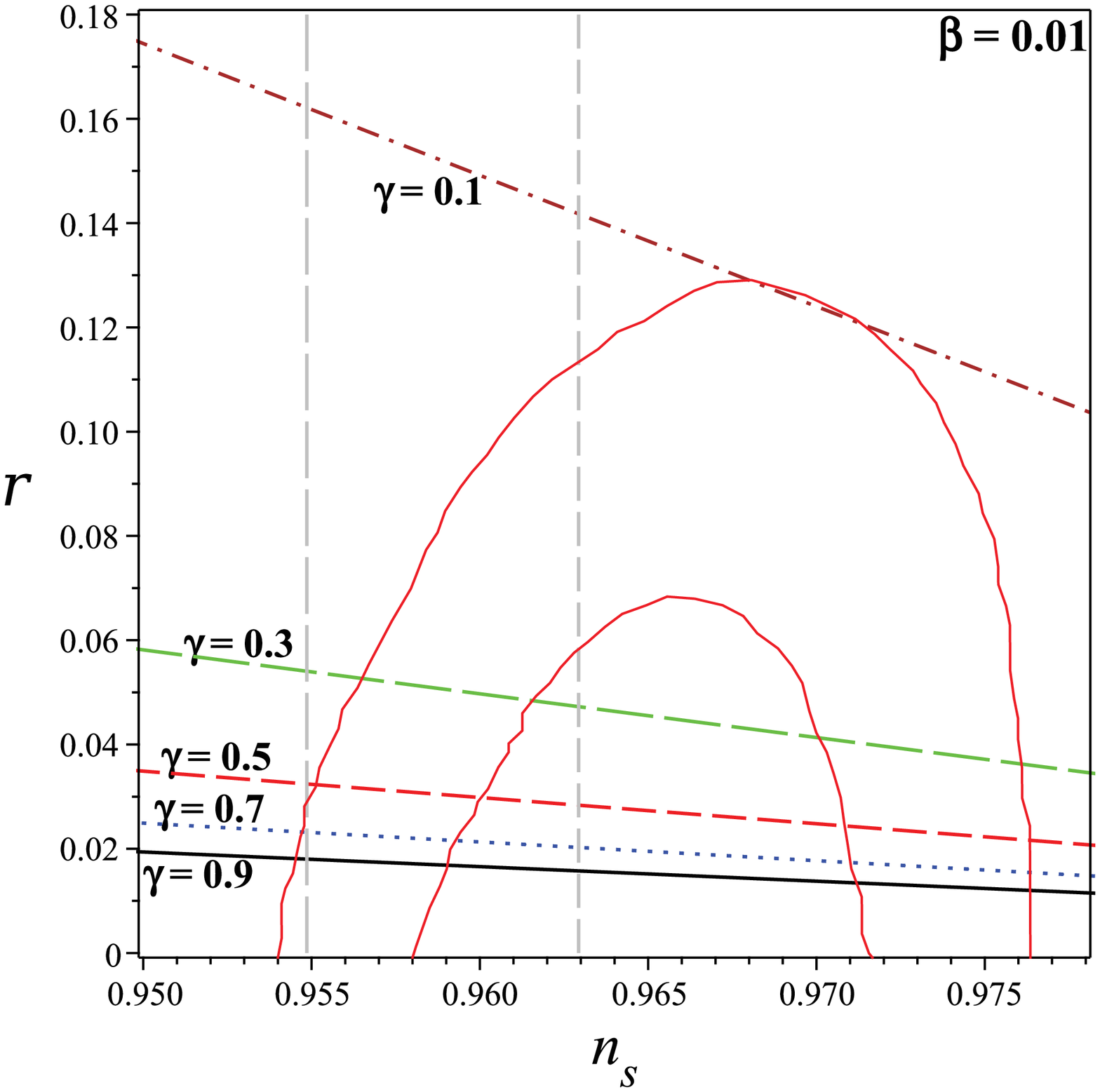}
\includegraphics[width=8cm]{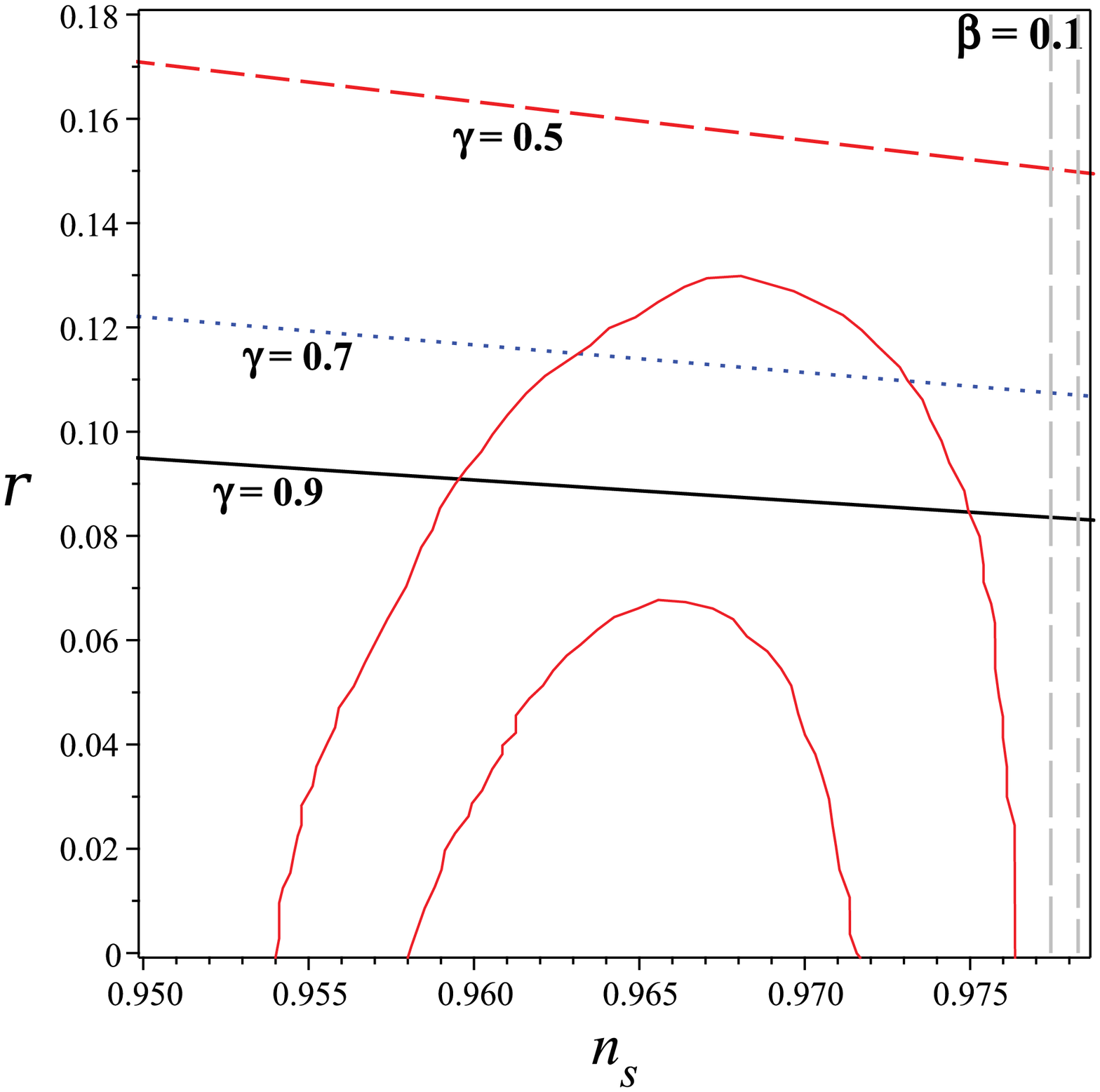}
\caption{The trajectories in $(r-n_s)$ plane for various values of $\gamma$ parameter. The left for $\beta=0.01$ and the right for $\beta=0.1$. The red contours show the results from Planck 2018. The vertical gray lines relate to $N=50$ (long-dashed) and $N=60$ (dashed).}\label{fig1}
\end{figure}

FIG.\ref{fig1} consists of two plots that show these dependencies in $(r-n_s)$ plane. The left one has plotted for the small value of $\beta=0.01$ that is closer to the condition of a slow-roll inflation than $\beta=0.1$ in the right one. We find that for $\beta=0.01$, all the trajectories for different values of $\gamma$ are generally in agreement with recent observations from Planck 2018, in the sense that they enter to the region $r<0.1$ when $n_s<1$, though the trajectories with greater $\gamma$ better fit the observations. To show this fact we have added two red contours into FIG.\ref{fig1} that have been obtained from Planck 2018 results and borrowed from \cite{Banerjee}. The smaller and the bigger contours are respectively assigned to 1$\sigma$ and 2$\sigma$ confidence regions. We understand that for $\beta=0.01$, the model under consideration is in good agreement with observations for $\gamma\lesssim0.1$, because respective trajectories pass through the contours. But, the case is different for $\beta=0.1$. With attention to the right plot one can realize that only the trajectories with large values of $\gamma$ satisfy the constraint on $r$, so that only the trajectories with $\gamma\gtrsim0.6$ go across the observational contours. Eventually, we conclude that the larger $\beta$, the greater values of $\gamma$ will fit the model with observations, so that for $\beta=0.1$, the trajectory with $\gamma=0.9$ does not even pass through the 1$\sigma$ confidence region.

Also, using Eqs.(\ref{eN}), (\ref{z}) and (\ref{ns}), one can realize that for each value of $\beta$, there is a one-to-one relationship between $n_s$ and $N$. For example, for the case $\beta=0.01$, we find $n_s\approx0.955$ and $n_s\approx0.963$ for $N=50$ and $N=60$, respectively. As the same way, for the case $\beta=0.1$, we find $n_s\approx0.977$ and $n_s\approx0.978$ for $N=50$ and $N=60$, respectively. The vertical gray lines in both the plots in FIG.\ref{fig1} relate to these values of $N$, so that in each of them the long-dashed line relates to $N=50$ while the dashed one relates to $N=60$. With attention to these lines which are the limits of uncertainty in the number of $e$-folds usually consider by Planck collaboration, it is once more clear that our model works well for smaller values of $\beta$.

\begin{figure}[h]
\centering
\includegraphics[width=8cm]{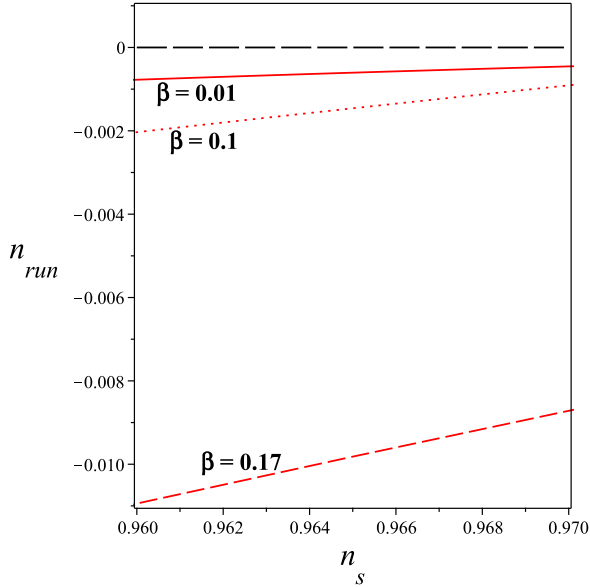}
\caption{The trajectories in $(n_{run}-n_s)$ plane for various values of $\beta$ parameter. In horizontal axis the 68\% confidence limits of $n_s$ have been considered. For all values of $\beta$, $n_{run}$ is negative.}\label{fig2}
\end{figure}

In \cite{Akrami} and \cite{Aghanim}, Planck 2018 collaboration has investigated the scale dependence of $n_s$. Their results reveal that running in the scalar spectral index has a very small negative value close to zero. Specifically, they have obtained $n_{run}=-0.0045\pm0.0067$ in 68\% confidence limit. FIG.\ref{fig2} demonstrates some possible trajectories in $(n_{run}-n_s)$ plane for different values of $\beta$ in our model. It is obvious that for all values of $\beta$, the running in the scalar spectral index is negative so that for each $\beta$, it increases when $n_s$ grows. Also, the greater $\beta$ leads to smaller values of $n_{run}$, generally. In FIG.\ref{fig2} we have considered the 68\% confidence limits of the scalar spectral index, $0.96\lesssim n_s\lesssim0.97$, from \cite{Aghanim}. We find that in this regime the model satisfies the 68\% confidence limit of $n_{run}$, up to $\beta\approx0.17$. Although if we take into account the number of $e$-folds, as we discussed above, the cases with $\beta<0.1$ are more acceptable. Therefore, one can realize that our model is in good agreement with observations for the smaller values of $\beta$ that is also compatible with the results in \cite{Motohashi2}.

\section{Conclusions}

In this manuscript we have investigated the cold inflationary Universe under the constant-roll conditions on the brane in DGP cosmology. In section 2, we have used a novel approach to solve the Friedman equation on the brane considering the constant-roll condition given by $\ddot{\phi}+\beta H\dot{\phi}=0$. The number of $e$-folds, $N$, has been used as the independent variable and the inflaton field, its potential energy and the Hubble parameter has been determined in terms of $N$.

In section 3, the evolution of primordial perturbations has been studied for both scalar and tensor perturbations. We have ignored the case $V_0\neq0$, because the analytical approach is only applicable for $V_0=0$. We have calculated some of the most important perturbation parameters such as the scalar spectral index $n_s$ and the tensor-to-scalar ratio $r$, in terms of model parameters. We found that irrespective of the induced parameter $\gamma$, the model under consideration coincides with observations for small values of the constant-roll parameter $\beta\lesssim0.02$. But for larger values of $\beta$, the constraint on $r$, will be close to observations only for greater values of $\gamma$. Studying the running in the scalar spectral index, we found that the cases with $\beta>0.17$ do not satisfy the 68\% confidence limits of $n_{run}$. Totally, our model shows good consistency with Planck 2018 results but for small values of $\beta$.

\section{Acknowledgments}

The authors would like to thank the referee for his/her valuable and helpful comments. This research is in memory of Prof. M. R. Setare who motivated us to study the constant-roll inflation.


\end{document}